\documentclass[12pt, natbib209]{emulateapj}
 \usepackage{psfrag}
 \shorttitle{Cepheid Mass Loss and Metallicity}
 \shortauthors{Neilson \& Lester}
  
\begin{document}

 \title{On the Enhancement of Mass Loss in Cepheids Due to Radial Pulsation. II. The Effect of Metallicity}
 \author{Hilding R. Neilson}
\affil{Department of Astronomy \& Astrophysics, University of Toronto, 50 St. George Street, Toronto, ON, Canada M5S 3H4}
\email{neilson@astro.utoronto.ca}
\author{John B. Lester}
\affil{University of Toronto Mississauga, Mississauga, ON, Canada L5L 1C6}

\begin{abstract}
It has been observed that Cepheids in the Magellanic Clouds have lower masses for the same luminosity than those in the Milky Way.  The model, from \cite{Neilson2008}, of pulsation--driven mass loss for Cepheids is applied to theoretical models of Cepheids with metallicity consistent with the Milky Way and Large and Small Magellanic Clouds.   The mass--loss model is analyzed using the metallicity correction of the Period--Luminosity relation to compare the ratio of mass loss of Cepheids with lower metallicity to that of Cepheids with solar metallicity.  It is determined that mass loss may be larger for the lower metallicity Cepheids,  counterintuitive to radiative driving estimates.  Also the mass--loss rates of theoretical Cepheid models are found to be up to $5\times 10^{-9}$ for Galactic Cepheids, $5\times 10^{-8}$ for Large Magellanic Cloud Cepheids, and  $2\times 10^{-7}M_\odot /yr$ for Small Magellanic Cloud Cepheids.  It is argued that mass loss increases as metallicity decreases for Cepheids with periods less than $20$ days and that mass loss decreases for longer periods.  Assuming dust forms in the wind of a Cepheid at some distance, the infrared excess of the models is computed, finding the infrared brightness is approximately a magnitude larger due to mass loss. The infrared magnitudes are compared to recently published Period--Luminosity relations as a test of our predictions. 
\end{abstract}

\section{Introduction}
The study of classical Cepheid variables has been a cornerstone of stellar and Galactic astrophysics since the discovery of the Period--Luminosity relation \citep{Leavitt1912}.  Observations of light curves and period provide information for testing stellar structure and evolution while the Cepheid Period--Luminosity relation makes them ideal standard candles.  It has been by understanding the physical structure and mechanism of pulsation that Cepheids have become powerful tools, but there are still phenomena being discovered.

One phenomenon was the recent discovery of infrared excesses from circumstellar material surrounding Galactic Cepheids \citep{Kervella2006,Merand2006,Merand2007}.  It has been suggested that the cause of the infrared excess is dust forming in a wind from the Cepheids \citep{Kervella2006}. There is additional evidence for mass loss, albeit circumstantial and wide ranging.    Analysis of IRAS observations \citep{McAlary1986, Deasy1988} infers mass--loss rates up to $10^{-6} M_\odot/yr$ for some Cepheids, although radio observations have placed upper limits of $10^{-9}$--$10^{-7}M_\odot/yr$ \citep{Welch1988}. Also \cite{Rodrigues1992} determined an upper limit of $\dot{M} \le 7\times 10^{-10}M_\odot/yr$ for SU Mus and \cite{Bohm-Vitense1994} found $\dot{M} \approx 5\times 10^{-5} M_\odot/yr$ for $l$ Car.  Furthermore the discovery of circumstellar material surrounding $l$ Car, Polaris, $\delta$ Cep, Y Oph \citep{Kervella2006, Merand2006, Merand2007} as well as RS Pup and SU Cas being associated with nebulae \citep{Havlen1972, Turner1984} provide more evidence for mass loss.  

Theoretical evidence for mass loss based on hydrodynamics is presented by \cite{Willson1986} and \cite{Brunish1989}. In \citet[hereafter Paper I]{Neilson2008} we developed an analytic method to calculate mass--loss rates for Cepheids in which the mass loss is driven by the combination of pulsation and radiation. The method was applied to observed Galactic Cepheids where the mass--loss rates are predicted to range from $10^{-10}$--$10^{-7}M_\odot/yr$. This implies that over the lifetime of Cepheid evolution, typically $10$ to $30 Myr$, a significant amount of mass can be lost.

Mass loss is interesting in Cepheids not only because it produces infrared excess and may affect the period of pulsation, but it has also been proposed as a solution to the Cepheid mass discrepancy, which is the difference between mass estimates of Cepheids by stellar evolution models and by pulsation models.  Historically, the mass determinations differed by as much as  $40\%$ \citep{Cox1980}.   The introduction of the OPAL opacities \citep{Rogers1992} brought a partial resolution to the mass discrepancy problem by reducing it to order $10\%$ \citep*{Moskalik1992}, however, the discrepancy is still an issue. \cite{Sebo1995} found a mass discrepancy of about $20\%$ for Cepheids with periods ranging from  $2.7$ to $30$ days in the Large Magallanic Cloud (LMC) cluster NGC 1850; likewise \cite*{Beaulieu2001} found similar results for Cepheids from the OGLE database for the LMC as well as a larger mass discrepancy for Small Magallanic Cloud (SMC) Cepheids.  \cite{Brocato2004} found masses of Cepheids in the LMC cluster NGC 1866 to be smaller than predicted by evolution by almost $30\%$ for short period Cepheids. The only dynamical mass estimates for Cepheids are from observations of Galactic Cepheids in binary systems; these are all found to have masses less than that predicted by evolutionary calculations \citep{Bohm-Vitense1998, Evans2006}.

The current status of the mass discrepancy for Galactic Cepheids was presented by \cite{Caputo2005}, who give the discrepancy as a function of period.  For short periods, the mass discrepancy is approximately $10\%$ -- $20\%$, and this decreases with increasing period.  \cite{Keller2008} argues against this conclusion based on an analysis of the behavior of the mass--luminosity relation of Cepheids at large masses, and he proposes the mass discrepancy is constant with respect to mass, about $20\%$.  The results differ because of the stellar evolution models used; \cite{Caputo2005} use stellar evolution models that do not include a prescription for mass loss on the main sequence or on the red giant branch while the \cite{Keller2008} results use stellar evolution models that do include mass loss.  Mass loss affects the evolution of the massive stars $M > 10M_\odot$ and causes a non--linear mass--luminosity relation for Cepheids,  the various mass--luminosity relations are shown in Figure 2 of \cite{Keller2008}.

\cite{Keller2002} explore the mass discrepancy of LMC Bump Cepheids using non--linear pulsation models and find a mass discrepancy of about $20\%$.  \cite{Keller2006} determined that the mass discrepancy for SMC Cepheids is about $20\%$ for Cepheids with mass $5$--$7M_\odot$, while the discrepancy for LMC Cepheids is about $17\%$, determined by matching MACHO and OGLE observations with non--linear pulsation models with varying mass and metallicity.  This observation has been verified for the LMC  \citep{Testa2007}, and it also agrees qualitatively with the analysis of \cite{Cordier2002}.   

While mass loss is one possible cause of the mass discrepancy, a second suggested source is convective core overshooting in the main sequence progenitors of Cepheids \citep{Huang1983}.  There is no direct evidence of this in Cepheids other than the mass discrepancy, coupled with the argument that mass loss is negligible over the lifetime of stars \citep{deJager1988, Girardi2000}.  However there is evidence for convective core overshooting from stellar isochrones fit to stellar cluster observation \citep[for example][]{Mucciarelli2007,Chiosi2007,Rosvick1998}. \cite*{Schroder1997} used observations of $\zeta$ Aurigae systems  and \cite*{Ribas2000} used observations of detached binaries to constrain convective core overshooting.  Representing convective core overshooting as $\Lambda_{\rm{CCO}} = \alpha H_{\rm{P}}$, where $H_{\rm{P}}$ is the pressure scale height, the free parameter $\alpha$ produces a match to the observations for $\alpha \approx 0.1$--$0.3$.  \cite{Ribas2000} find larger convective core overshooting for the binaries V380 Cyg, and HV 2274.  This evidence is circumstantial because the parameter describing convective core overshoot is obtained by matching models that  might ignore other physical processes.

The purpose of this article is to investigate the effect of metallicity on mass loss in Cepheids using the analytic model from Paper I.  In particular, we wish to explore the potential for Cepheid mass loss in the Large and Small Magellanic Clouds.  Understanding Cepheid mass loss in the Clouds will provide insight into the source of the mass discrepancy. The goals of this work are to differentiate how 
mass loss and convective core overshoot would contribute to the mass discrepancy, as well as to predict observational consequences of mass loss, such as infrared excess.  We start this exploration by deriving the metallicity dependence of the analytical mass--loss formulation from Paper I  in Section 2.  In Section 3, the analytical model is applied to Cepheids in the LMC and SMC to predict the potential behavior at lower metallicity.  Mass--loss rates are calculated in Section 4 for theoretical models of Galactic, LMC and SMC Cepheids and the resulting infrared luminosities  are predicted in Section 5. The dependence of the mass discrepancy on pulsation--driven mass loss is explored in Section 6.

\section{The Metallicity Dependence of the Analytical Mass Loss Relation}
In Paper I, an analytic solution for the mass--loss rate as a function of pulsation phase was derived based on an extension of the method of \cite*{Castor1975}.  This method solves the momentum equation which balances the force of gravity with the forces due to continuum and line radiative driving and contributions due to pulsation and shocks generated by the pulsation.  An approximate solution to the momentum equation is derived in Paper I for the mass--loss rate at some phase of pulsation,
\begin{eqnarray}\label{e1}
&& \nonumber \dot{M} = \left[\frac{\sigma_e L k}{4\pi c} \frac{Z}{Z_\odot}(1- \alpha)\right]^{1/\alpha}\times \\
&& \left(\frac{4\pi}{\sigma_ev_{th}}\right)\left(\frac{\alpha}{1-\alpha}\right)[GM(1-\Gamma_e) -\zeta R^2]^{1-1/\alpha}.
\end{eqnarray}
The mass--loss rate is an implicit function of the pulsation phase and the integral of Equation \ref{e1} with respect to phase over the period is the phase--averaged mass--loss rate.  The variables $k$ and $\alpha$ are force multiplier parameters that depend on the gravity and effective temperature of a star,  computed from models of stellar atmospheres. For Cepheids, the values used are $k = 0.064$ and $\alpha = 0.465$ \citep{Abbott1982}.  The luminosity, mass and radius are denoted  $L$, $M$, and $R$ respectively while $\sigma_e$ is the electron scattering opacity, and $v_{th}$ is the thermal velocity.  The function $\Gamma_e \equiv \sigma_eL/(4\pi cGM)$ is the ratio of the acceleration due to continuum driving and gravity, while the function $\zeta$ is the sum of the acceleration due to pulsation and shocks,
\begin{eqnarray}\label{e2}
&&  \nonumber\zeta  = \Delta R \omega^2 \cos(\omega t) + P^{-1}\frac{du_\delta }{d\phi}   \left( \frac{L}{L_\delta}\right)^{1/8} \\
&& \times\left(\frac{P}{P_\delta}\right)^{-7/24}\left(\frac{M}{M_\delta}\right)^{5/12}.
\end{eqnarray}
The first term of Equation \ref{e2} represents the acceleration due to pulsation where $\Delta R$ is the amplitude of radius variation, $\omega = 2\pi/P $ is the angular frequency of pulsation and $t$ is the time.  The second term of $\zeta$ is the acceleration due to shocks, where $P $ is the period of pulsation, $du_\delta/d\phi$ is the change of velocity with respect to phase of the shocks at the surface for a model of the atmosphere of $\delta$ Cep \citep{Fokin1996}, $L_\delta$, $P_\delta$, and $M_\delta$ are the luminosity, pulsation period and mass of the model of $\delta$ Cep, and $L $ and $M $ is the luminosity and mass of the Cepheid. The function $\zeta$ and shock acceleration are derived in Paper I.

This approximate solution for the mass--loss rate of a pulsating star can be used to evaluate the dependence on metallicity.  Consider first the dependence of mass loss on metallicity for winds driven by radiation alone.  In this case the mass--loss rate is given by Equation \ref{e1} with $\zeta = 0$ at all phases of pulsation.  For Cepheids evolving on the second crossing of the instability strip the luminosity is proportional to the mass, with the same dependence for different metallicities \citep{Castellani1992, Girardi2000}.  The instability strip is somewhat hotter for smaller metallicities \citep{Bono2000}, but there is significant overlap for the metallicities in question.  Therefore the temperature range does not contribute significantly.  If we consider two Cepheids with differing metallicities but same mass and luminosity then from Equation \ref{e1} with $\zeta = 0$ the comparison of the mass--loss rates for two metallicities $Z_1$ and $Z_2$ is 
\begin{equation}\label{e3}
\frac{\dot{M}_1}{\dot{M}_2} = \left(\frac{Z_1}{Z_2}\right)^{1/\alpha}.
\end{equation}
This implies that Cepheids in the LMC and SMC would have lower mass--loss rates then those in the Milky Way as the relative metallicity for the LMC is $0.4$ and for the SMC is $0.2$.  Because $\alpha = 0.465$, $\dot{M}_{\rm{LMC}}/\dot{M}_{\rm{MW}} = 0.139$ and   $\dot{M}_{\rm{SMC}}/\dot{M}_{\rm{MW}} = 0.031$. Thus the radiative--driven mass--loss rates in the LMC and SMC are very small relative to the Milky Way. However the metallicity is important in the calculation of pulsation--driven mass--loss rates as well as radiative--driven mass--loss rates.  If we again consider Equation \ref{e1} to describe the ratio of mass--loss rates for two different metallicities then
\begin{equation}\label{e4}
\frac{\dot{M}_1}{\dot{M}_2} = \left(\frac{Z_1}{Z_2}\right)^{1/\alpha}\left[\frac{GM_{1}(1-\Gamma_{e,1}) -\zeta_1 R_{ 1}^2}{GM_{ 2}(1-\Gamma_{e,2}) -\zeta_2 R_{ 2}^2}\right]^{1-1/\alpha}.
\end{equation}
This can be rewritten as
\begin{eqnarray}\label{e5}
&& \nonumber\frac{\dot{M}_1}{\dot{M}_2} = \left(\frac{Z_1}{Z_2}\right)^{1/\alpha}\left[\left(\frac{g_{\rm{eff},1}}{g_{\rm{eff},2}}\right)\right. \times \\
&& \left.\left(\frac{1 - a_{\rm{puls},1}/g_{\rm{eff},1} - a_{\rm{Shock}, 1}/g_{\rm{eff},1}}{1 -  a_{\rm{puls},2}/g_{\rm{eff},2} - a_{\rm{Shock}, 2}/g_{\rm{eff},2} }\right)\right]^{1-1/\alpha},
\end{eqnarray}
where the terms $a_{\rm{Shock}}$ and $a_p$ are the shock and the pulsation acceleration terms of the function $\zeta$ respectively, and $g_{\rm{eff}}$ is the effective gravity, $GM(1-\Gamma_e)$. The ratio of the shock acceleration to the effective gravity is
\begin{eqnarray}\label{e6}
&& \nonumber\frac{a_{\rm{Shock}}}{g_{\rm{eff}}} =  \frac{du_\delta}{Pd\phi}    \left( \frac{L}{L_\delta}\right)^{1/8} \left(\frac{P}{P_\delta}\right)^{-7/24}\left(\frac{M}{M_\delta}\right)^{5/12}\\
&&\times \left[\frac{R^2}{GM(1-\Gamma_e)}\right] = C_0P^{-1.29}M^{-0.58}R^2,
\end{eqnarray}
and the ratio of the pulsation acceleration and the effective gravity is
\begin{equation}\label{e7}
\frac{a_{\rm{puls}}}{g_{\rm{eff}}} = \frac{2\pi \Delta R R^2}{GM(1 - \Gamma_e)}.
\end{equation}

The ratio of the analytic mass--loss rates can be used to test the behavior of a Magellanic Cloud Cepheid relative to a Galactic Cepheid, and to investigate if pulsation--driven mass loss is significant for lower metallicities. To do this, Equation \ref{e5} must be rewritten such that all variables are in relative non--dimensional units. Consider two Cepheids, labeled 1 and 2, and express the ratio of the pulsation acceleration for the Cepheid number 2 be in terms of the effective gravity, called $a_p$.  Second, write the balance of force for Cepheid number 2 as the non--dimensional parameter $F = 1 -  a_{p} - a_{\rm{Shock}, 2}/g_{\rm{eff},2}$. Also it is assumed that $\Gamma_e \approx 0$ and can be ignored in the calculation because the electron number is small for stars with effective temperatures ranging from  $4000$ to $6000K$. The small electron number causes an acceleration that is small relative to the gravity.   The acceleration due to pulsation for Cepheid 1 relative to Cepheid 2 is 
\begin{equation} \label{e8}
\frac{a_{\rm{puls},1}}{g_{\rm{eff},1}}  = a_p \left(\frac{\Delta R_1}{\Delta R_2}\right)\left(\frac{R_{1}}{R_{ 2}}\right)^2\left(\frac{P_{ 1}}{P_{ 2}}\right)^{-2}\left(\frac{M_{ 1}}{M_{ 2}}\right)^{-1},
\end{equation}
and the ratio of the shock acceleration and the effective gravity is 
\begin{eqnarray}\label{e9}
&& \nonumber\frac{a_{\rm{Shock},1}}{g_{\rm{eff},1}} = (1 -  a_p - F)\left(\frac{P_{ 1}}{P_{ 2}}\right)^{-1.29}\\   
&&\times \left(\frac{M_{ 1}}{M_{ 2}}\right)^{-0.58}\left(\frac{R_{ 1}}{R_{ 2}}\right)^2.
\end{eqnarray}
Substituting these results into Equation \ref{e5} and writing all quantities with a subscript of $1$ in units of the quantities denoted with a subscript of $2$, the relative mass--loss rate of Cepheid $1$ is
\begin{eqnarray}\label{e10}
&&\nonumber \dot{M}_1(\dot{M}_2) = Z_1^{1/\alpha} \left\{\left(\frac{M_{ 1}}{F}\right) \left(1  -a_p\Delta R_1R_{ 1}^2P_{ 1}^{-2}M_{ 1}^{-1} \right. \right. \\
&&\left.\left.- (1-a_p-F)P_{ 1}^{-1.29}M_{ 1}^{-0.58}R_{ 1}^2\right)\right\}^{1-1/\alpha}.
\end{eqnarray}
 Equation \ref{e10} can be used to probe the mass--loss rates for low metallicity Cepheids as a function of relative period with the following free parameters: $a_p$, $F$, $M_1$, $R_1$,  and $\Delta R_1$. 

If the metallicity $Z_1$ is less than unity then the relative mass--loss rates, $\dot{M}_1$, tend to be smaller based on the explicit dependence of metallicity. However, the ratio of mass--loss rates may be greater than or equal to unity if the term inside the curly brackets is significantly less than one. This term is significantly less than one only when
\begin{eqnarray}\label{e11}
&& \nonumber 1-a_p\Delta R_1R_{ 1}^2P_{ 1}^{-2}M_{ 1}^{-1} \\
&&- (1-a_p-F)P_{ 1}^{-1.29}M_{ 1}^{-0.58}R_{ 1}^2 <<  1,
\end{eqnarray}
and furthermore this term must be smaller than the balance of forces, $F$. There are many ways for this function to be small, one of which is the relative period.  If $P_1 <1$ and $R_1$ and $M_1$ are approximately unity, then this function is less than one; however how much less than one $P_1$ needs to be will depend on the values of $a_p$ and $F$.  The value of $a_p$, the ratio of the acceleration due to pulsation to the effective gravity, is generally in the range $0$ to $0.1$ for Galactic Cepheids. Therefore $a_p$ plays only a small role in affecting the mass--loss rate, implying that the shock acceleration term is the dominate term, which means the parameter $F$ needs to be small to satisfy Equation \ref{e11}.  A small value of $F$ suggests that the balance of forces for the reference Galactic Cepheid is small and the mass--loss rate is large.
\begin{figure*}[t]
\plotone{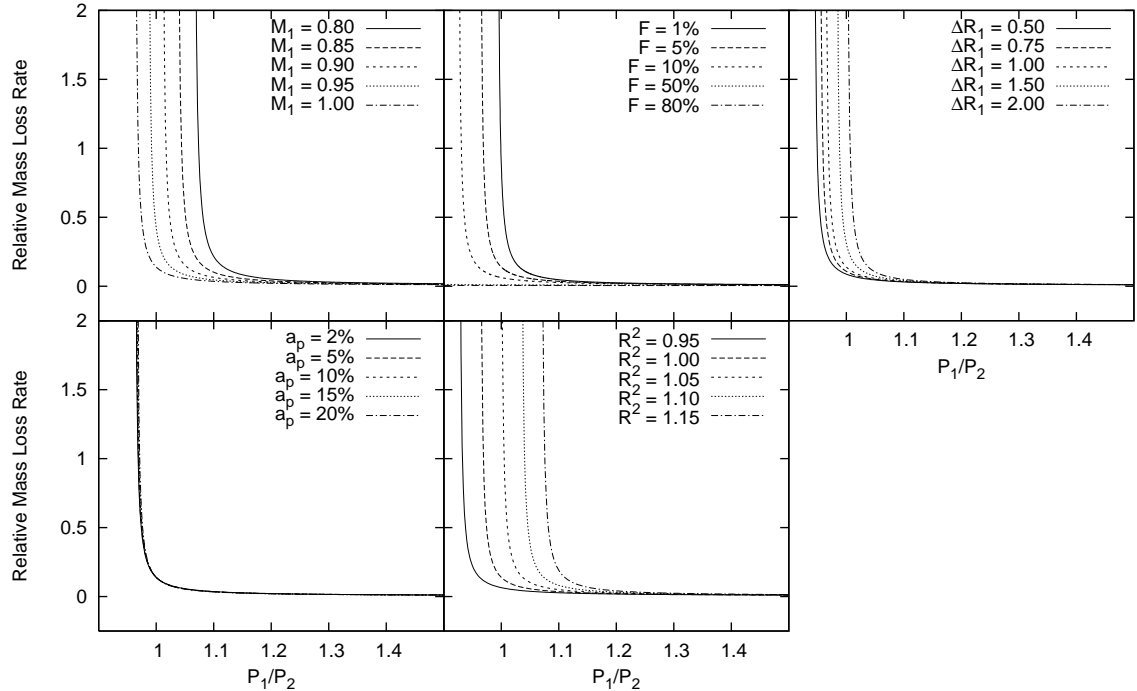}
\caption{The mass--loss rate of a LMC Cepheid relative to a similar Cepheid with solar metallicity as a function of period and the effect of varying the other parameters. When the parameters are held constant, they are $R_1=1$, $M_1=1$, $\Delta R_1=1$, $F = 5\%$, and $a_p = 5\%$.}
\label{f1}
\end{figure*}

The inequality given by Equation \ref{e11} is satisfied when the relative radius is greater than unity or when the mass is less than unity, but this also requires that the value of $F$ be small.  Varying the amplitude of radius pulsation affects the result as well; if the amplitude is greater than unity for lower metallicity then the pulsation terms increases linearly and thus decreases Equation \ref{e11}.  The result of varying the values of $a_p$, $F$, $R_1^2$, $M_1$, and $\Delta R_1$ in Equation \ref{e10} gives the mass--loss rate of Cepheid $1$, $\dot{M}_1$, shown in Figure \ref{f1} as a function of $P_1$ for the relative metallicity of the LMC to the Milky Way.  From the plots, it is clear that the relative mass--loss rate can become very large even at lower metallicity.  Applying the same analysis for the mass loss of SMC Cepheids relative to Galactic Cepheids the curves shows similar behavior.  The effect of the metallicity in the analysis is to shift the curve up and down along the mass--loss rate axis. For a relative mass--loss rate of unity the relative period is approximately the same for both Cepheids with LMC metallicity and Cepheids with SMC metallicity since it is in the nonlinear regime given by the term in the curly brackets in Equation \ref{e10}.

The mass--loss rates are larger at lower metallicity if the shocks are more efficient.  These shocks are generated in the hydrogen and helium partial ionization zones of Cepheids and at lower metallicity the partial ionization zones produce more energy when the layers become ionized.  This is because of the larger fractional mass hydrogen and helium. Therefore the shocks have more momentum to input into the wind.  The lower metallicity and possibly larger abundance of helium also affect the global properties of Cepheids \citep{Marconi2005}, such as the pulsation period, and these metallicity--dependent global properties appear in the shock acceleration formula making it implicitly metallicity dependent.

\section{Comparison of the Analytic Predictions with Observations}
It has been argued that pulsation--driven mass loss generally decreases for lower metallicities, but there are combinations of parameters that predict larger relative mass--loss rates. One might ask what the parameters describing Magellanic Cloud Cepheids would predict.  

The average relative period of the LMC and SMC Cepheids can be determined using the metallicity correction. The metallicity correction is a constant that is added to the Period--Luminosity relation to account for the fact that Cepheids with different metallicities have different luminosities for the same pulsation period.  The Hubble Key Project on the Extragalactic Distance Scale \citep{Kennicutt1998} used the Period--Luminosity relation to determine the distance to galaxies in the local group, but it required a correction for the metallicity of approximately $\delta(m - M)/\delta [M/H] = -0.25$ mag/dex.  This correction factor has been verified by other studies \citep{Tammann2003, Sakai2004, Groenewegen2004, Romaniello2005, Groenewegen2003, Gieren2005} but the exact value is uncertain.  For instance, fecent theoretical studies using non--linear hydrodynamic models find the period--luminosity relation depends significantly on the metallicity as well as on the helium abundance \citep{Fiorentino2002, Marconi2005}.  \cite{Sasselov1997} argued that the metallicity correction, $\delta \mu$, could be written as a function of $Z$,
\begin{equation}\label{e12}
\delta \mu = 0.44\log\left(\frac{Z}{Z_{\rm{LMC}}}\right).
\end{equation}
Combining this with the period--luminosity relation for the LMC,
\begin{equation}\label{e13}
M_V = -2.760(\log P -1) - 4.218,
\end{equation}
\citep{Freedman2001}, one finds the period ratio for the LMC and SMC relative to Milky Way Cepheids with the same luminosity to be $P_{\rm{LMC}}/P_{\rm{MW}} \sim 1.16$ and $P_{\rm{SMC}}/P_{\rm{MW}} \sim 1.30$.

The relative mass of Cepheids in the LMC and SMC can be estimated by comparing the mass discrepancy in the Magellanic Clouds and the Milky Way \citep{Keller2006}.  The mass discrepancy is a function of metallicity implying that for the same luminosity the mass of Cepheids is a function of metallicity.  Therefore the mass of SMC Cepheids may be up to approximately $3\%$ smaller than LMC Cepheids, which in turn is about $3\%$ smaller than Galactic Cepheids. 

\begin{figure*}[t]
\epsscale{1.15}
\plottwo{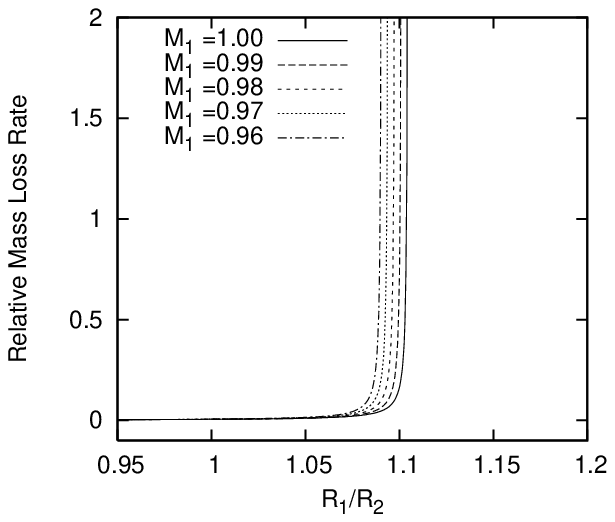}{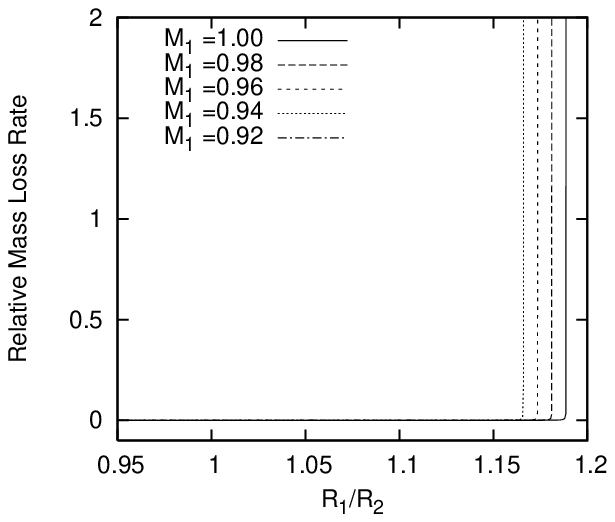}
\caption{The relative mass--loss rates for LMC (Left Panel) and SMC (Right Panel) Cepheid as a function of relative radius with period ratios $P_{\rm{LMC}}/P_{\rm{MW}} = 1.16$ and $P_{\rm{SMC}}/P_{\rm{MW}} = 1.3$ respectively.}
\label{f2}
\end{figure*}

The values of the relative radius and relative amplitude of radius variation are difficult to approximate. Radial velocity observations suggest that the radii of Cepheids in the Magellanic Clouds follow a similar Period--Radius relation as Galactic Cepheids \citep{Storm2004b,Storm2005}.  Therefore varying these parameters can provide a test of mass loss with the period ratios quoted for the LMC and SMC.  Figure \ref{f2} shows the relative mass--loss rate for LMC (Left Panel) and SMC (Right Panel) Cepheids as a function of the mean radius for corresponding relative period.  The analysis assumes the value of $F$ is $0.01$, $\Delta R_1 = 1.15$, and $a_p  = 0.05$.  Each curve represents the possible relative mass range suggested by mass discrepancy studies, $0.96$ to $1$ for the LMC and $0.92$ to $1$ for the SMC.  The results are insensitive to small variations of $F$, $\Delta R_1$, and $a_p$, implying that mass loss in the Clouds is significant if the radii of Cepheids are about $10\%$ and $16\%$ larger in the LMC and SMC respectively than Galactic Cepheids with the same luminosity.  This may be tested by measuring the angular diameters of Cepheids using interferometry, \cite{Mourard2008} argued that it is possible to measure these angular diameters with the Very Large Telescope Interferometer using differential interferometry techniques.

The plausibility of this required larger radius may be checked by considering the observed Period--Radius (PR) relation \citep{Gieren1999, Groenewegen2007}. The PR relation for Galactic, LMC and SMC Cepheids is 
\begin{equation}
  \log R = 0.68 \log P + 1.146,
\end{equation}
and for the period ratios of $1.16$ and $1.3$ the relative LMC and SMC Cepheid radius is $1.1$ and $1.2$ respectively.  This implies that it is possible that LMC and SMC Cepheids have larger mass--loss rates than Galactic Cepheids when the mass loss of the Galactic Cepheid is significant.

The metallicity correction that is determined using theoretical models suggests that metal--poor Cepheids are more luminous for the same period.  In other words, the periods of metal--poor Cepheids are shorter for the same luminosity \citep{Fiorentino2002}.  In this case the value of $P_1$ is less than unity. The relative mass--loss rate will be greater than unity for LMC and SMC Cepheids as long as the other parameters are $R_1 \ge 1$, $\Delta R_1 \ge 1$ and $M_1 \le 1$.

This analysis has shown that it is possible for metal--poor Cepheids to have significant mass loss, similar to and even greater than the mass--loss rates for Galactic Cepheids for both relative period regimes suggested by theoretical and observed values of the metallicity correction of the Period--Luminosity relation.

\section{Predicting Mass--Loss Rates For Theoretical Model Cepheids}
The previous analysis is suggestive, but it can be carried further to determine, quantitatively, how much larger the mass--loss rates are for LMC and SMC Cepheids, at least in the regime where the periods of LMC and SMC Cepheids are smaller than those of Milky Way Cepheids for the same luminosity.  To compute the mass--loss rate for a Cepheid, it is necessary to know the following parameters that describe the Cepheid: the mass, luminosity, radius, and pulsation period, as well as the amplitudes of the variation of the luminosity and radius.  While these quantities have been determined for many Galactic Cepheids \citep{Moskalik2005}, this has not been done for LMC and SMC Cepheids.  Therefore theoretical models of Cepheids from \cite{Bono2000} are used to predict mass--loss rates.  The method of solving the mass loss in the pulsating case is derived in Paper I.

The models include all the necessary information to calculate pulsation--driven mass--loss rates using the enhanced CAK method.  There are two types of models describing the mass--luminosity relations: one is based on canonical stellar evolution models, and the second assumes convective core overshoot, which results in a larger luminosity for the same mass as the canonical models.  This does not mean the models necessarily represent only convective core overshoot; for instance the steeper mass--luminosity relation could also represent mass loss during stages of evolution before the second crossing of the instability strip. The masses of the models are $5, 7, 9$ and $11M_\odot$ and span the temperature range of the instability strip.

\begin{figure*}[t]
\epsscale{1.1}
\plottwo{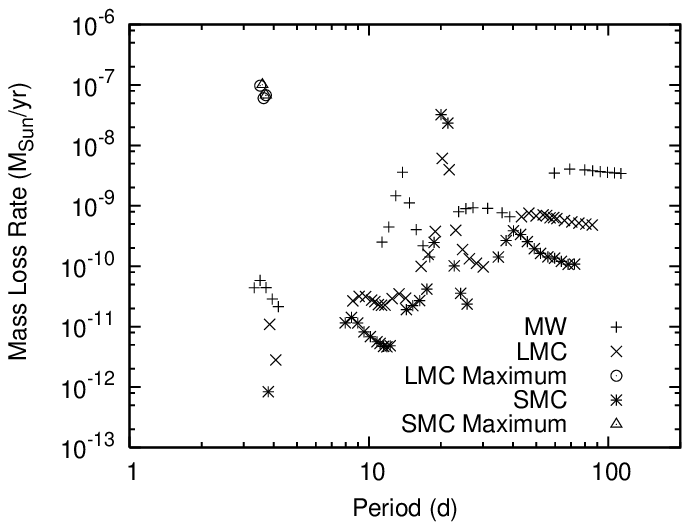}{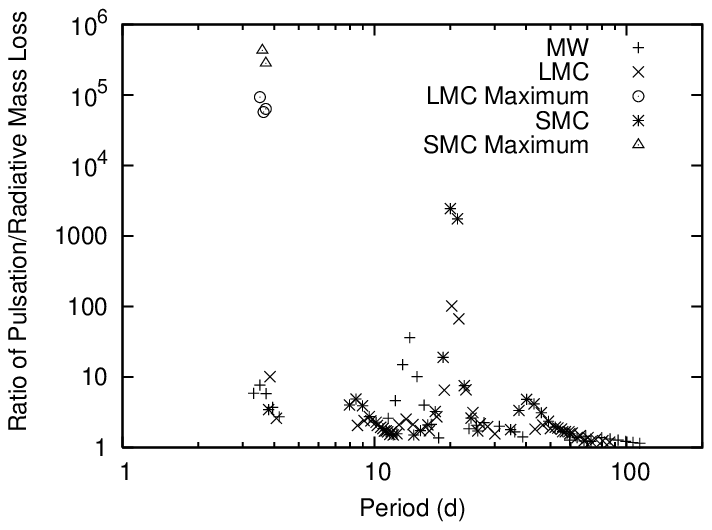}
\caption{(Left Panel) The predicted mass--loss rates of the theoretical Cepheid models with a canonical mass--luminosity relation \citep{Bono2000}. (Right Panel) The enhancement of the mass loss as given by the ratio of the pulsation--driven mass--loss rate to the radiative--driven mass--loss rate that ignores pulsation effects. The points given in the key as LMC/SMC Maximum are explained in the text.}
\label{f3}
\end{figure*}

The mass--loss rates for model Cepheids in the Milky Way, LMC and SMC are shown in Figure \ref{f3} as a function of pulsation period for the canonical models, along with the ratio of the pulsation mass--loss rates and the radiative--driven mass--loss rates.  The ratio, the enhancement of mass loss, is a measure of the dependence of the wind on the pulsation plus shock terms in the mass--loss calculation. The points labeled as LMC or SMC Maximum are based on model Cepheids where the pulsation period is so small that the sum of the acceleration due to pulsation and shocks is greater than the effective gravity of the Cepheid.  In this limit the calculation of the mass--loss rate becomes imaginary.  These points are calculated by using the same parameters from the model Cepheid but increasing the period by steps of $0.01$ day until the calculation is stable.  The physical justification for this is based on how the rate of change of period depends on mass loss.  It was shown in Paper I that mass loss acts to increase the rate of period change, making it more positive.  For most Cepheids, the contribution to the period change due to mass loss is small compared to the contribution due to evolution, but for Cepheids near the blue edge of the instability strip the effect of evolution on period change decreases, making the contribution of mass loss more significant.  The period is also smaller, further increasing the mass--loss rate and the effect of mass loss on the period change.
 
\begin{figure*}[t]
\epsscale{1.1}
\plottwo{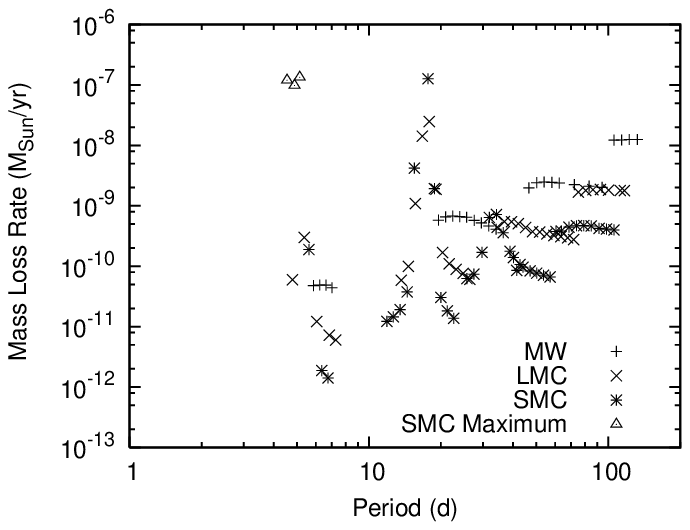}{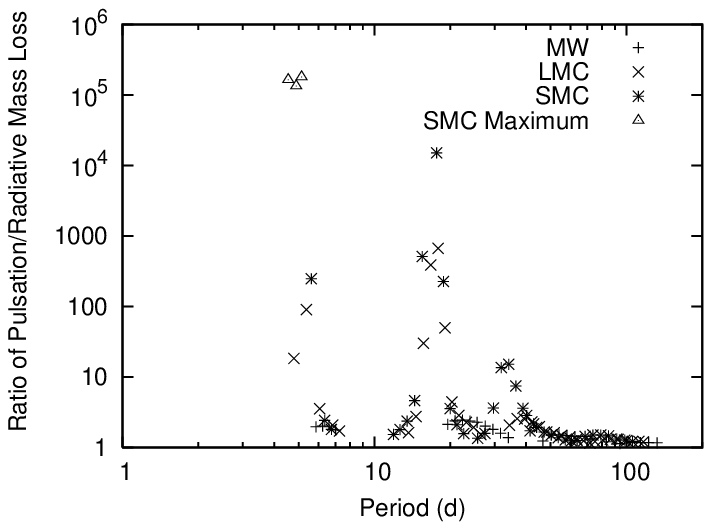}
\caption{Same as Figure \ref{f3} but for Cepheid models assuming a convective overshoot mass--luminosity relation.}
\label{f4}
\end{figure*}

\begin{figure*}[t]
\epsscale{1.1}
\plottwo{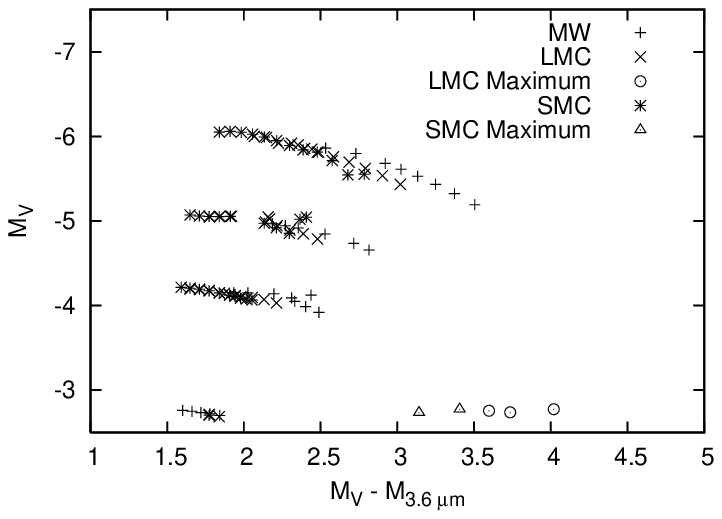}{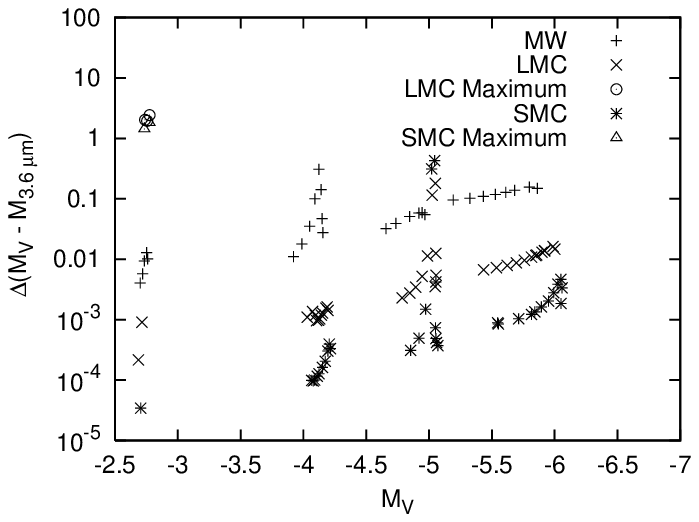}
\caption{(Left Panel) The color--magnitude plot of the model Cepheids based on the canonical mass--luminosity relation.  Cepheids with large mass--loss rates have significant color excess. The different sequences are for different masses, the brightest Cepheids are the most massive. (Right Panel) The color excess due to dust in the circumstellar shells.}
\label{f5}
\end{figure*}
\begin{figure*}[t]
\epsscale{1.1}
\plottwo{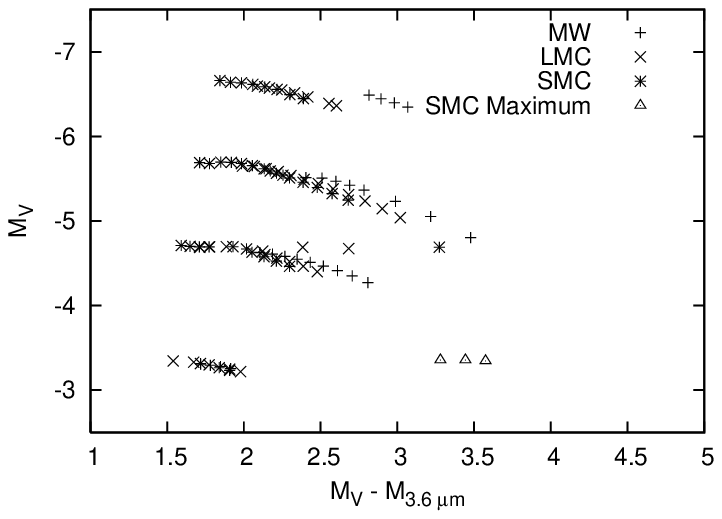}{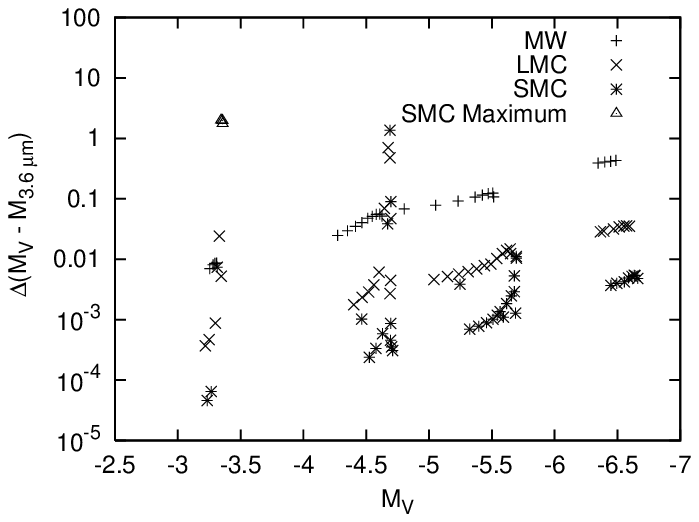}
\caption{Same a Figure \ref{f5} but for the model Cepheids with the convective overshooting mass--luminosity relation.}
\label{f6}
\end{figure*}

\begin{figure*}[t]
 \plottwo{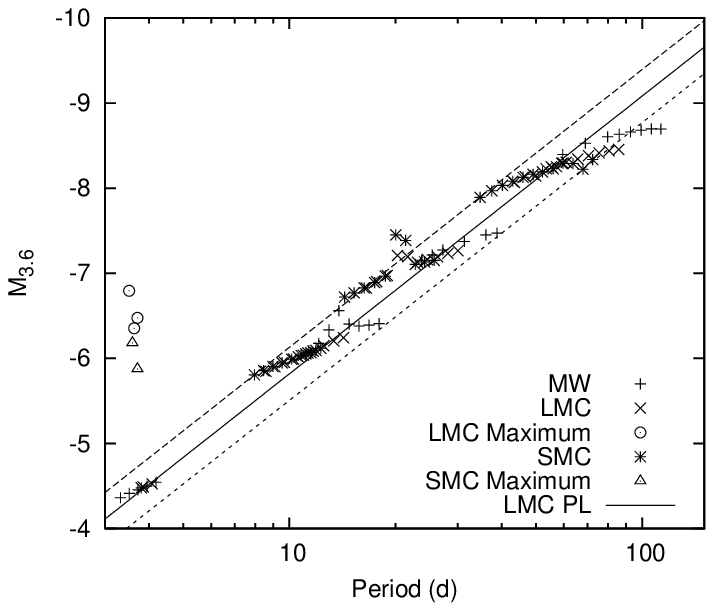}{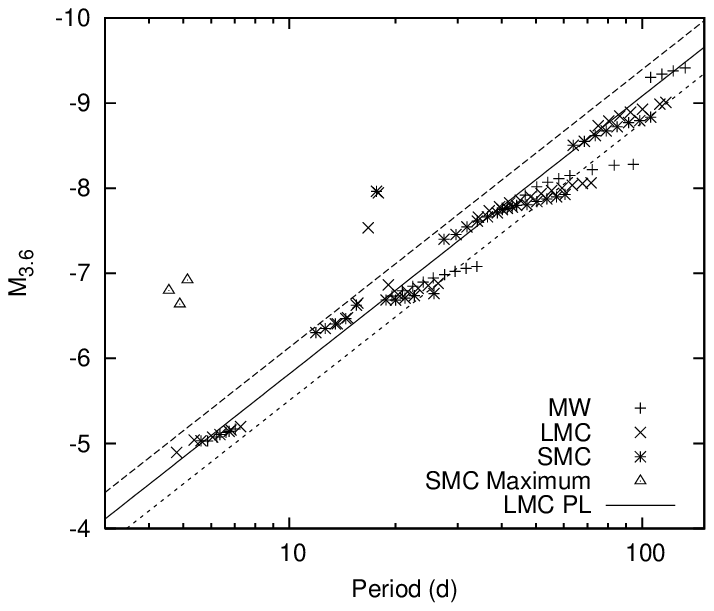}
 \caption{The Luminosity at $3.6$ $\mu m$ of the theoretical Cepheid models from \cite{Bono2000} including the effect of mass loss for the three values of the metallicity.  The models with a canonical mass--luminosity relation is shown in the Left Panel while the convective core overshooting models are shown in the Right Panel.  The LMC Period--Luminosity relation from \cite{Ngeow2008} with the $3\sigma$ deviation is shown for comparison.}
 \label{f7}
 \end{figure*}

The mass--loss rates for the SMC models shown in Figure \ref{f3} are larger than that for LMC and Milky Way at periods about $3.5$ days, denoted by LMC and SMC Maximum in the figures.  This is also true at   periods near $20$ days. These periods correspond to masses of $5$ and $9 M_\odot$ respectively.  The increase of mass--loss rates with decreasing metallicity is clear in these two cases, but for the other models the effect is not as obvious.  It is interesting that the long period Cepheids have lower mass--loss rates with decreasing metallicity.  This is consistent with the observation that the mass discrepancy of Cepheids decrease with increasing period for the Milky Way \citep{Caputo2005}, and it suggests similar behavior for LMC and SMC Cepheids.   The mass--loss enhancement, in the right panel of Figure \ref{f3}, shows that the mass loss in the SMC Cepheids is more sensitive to the effects of pulsation for Cepheids in the LMC and Milky Way; for Cepheids with periods near $10$ days the SMC mass loss is enhanced by a factor of 10 relative to the LMC.  This verifies the the argument that when mass loss is enhanced by pulsation for Galactic Cepheids, the mass loss is further amplified for similar Cepheids at the same position on the instability strip but with lower metallicities.  

The analysis is repeated with the steeper mass--luminosity relation associated with convective core overshoot and is shown in Figure \ref{f4}.  The dependence of pulsation--driven mass loss on metallicity is more clear in this case.  The mass--loss rates for SMC Cepheids may be larger than those of the LMC and the Milky Way for periods up to about $20$ days which is consistent with the result shown in Figure \ref{f3}.  For periods greater than $20$ days, the mass--loss rate decreases as a function of metallicity where pulsation and shocks are not as effective.  The mass--loss enhancement, shown in the right panel of Figure \ref{f4}, implies pulsation is more efficient for driving mass loss as metallicity decreases.  The mass--loss rates are also largely independent of the mass--luminosity relation for the theoretical Cepheid models.

\section{Predictions of Infrared Excess Due to Mass Loss}
It has been hypothesized that mass loss is important in LMC and SMC Cepheids.  In Paper I,  the circumstellar shells that have been observed as infrared excess in Galactic Cepheids were modeled as dust forming in the winds of Cepheids.  This luminosity depends on the fraction of the total amount of gas that forms dust, the dust--to--gas ratio, which for the Milky Way is approximately $1/100$.  This quantity is dependent on the composition of the gas, meaning the dust--to--gas ratio is smaller for the lower metallicity of the LMC and SMC.  For this work the dust--to--gas ratio is assumed to scale linearly with the relative metallicity, meaning that the ratio for the LMC is $1/250$ and SMC is  $1/500$.  However these are maximum values for the dust--to--gas ratios; for instance \cite{Clayton1985} found the LMC dust--to--gas ratio is about $1/400$ and \cite{Weingartner2001} determined that the ratio is about $1/1000$. These chosen dust--to--gas values are the maximum values but are uncertain by only about a factor of 5.

If mass loss is important in LMC and SMC Cepheids then the winds should produce observable infrared excess.  A Color--Magnitude Diagram, $M_V$ versus $M_V - M_{3.6\mu m}$, is shown in Figure \ref{f5} for the Cepheid models based on the canonical mass--luminosity relation.  Along with that plot, the difference of the colors due to the star plus circumstellar shell relative to the star alone is also shown in the right panel.   Because the luminosity of a circumstellar shell is due to dust, it will have minimal effect on the value $M_V$, but the dust contributes significantly to the $3.6$ $\mu m$ luminosity.  Therefore color excess increases if the mass--loss rate is large.  

The color--excess analysis is repeated for the convective core overshoot models and is shown in Figure \ref{f6}. In this case, the color excess is more pronounced though for only a few points. In both cases, the color excess due to circumstellar shells is significant.  If mass loss is important in LMC and SMC Cepheids then observations of a large population of Cepheids should be able test this model.
 
Infrared excess may also affect the infrared Period--Luminosity relation. The Period--Luminosity relation for the LMC was recently presented by both \cite{Ngeow2008} and \cite{Freedman2008} in IRAC bands using SAGE data \citep{Meixner2006}.  The PL relation was also determined in J, H and K--bands by \cite{Persson2004} for a sample of 92 Cepheids.  In \cite{Ngeow2008} and \cite{Persson2004} blending is an important issue because of the lower resolution of the infrared observations; while \cite{Freedman2008} avoids the issue by using a subsample from \cite{Persson2004}.  Blending acts to make the Cepheid appear more luminous in the infrared, just as mass loss would do, making the two difficult to distinguish from photometry.  In Figure \ref{f7}, the luminosities of the theoretical Cepheids, including infrared excess due to mass loss are plotted as a function of period for the Galactic, LMC and SMC sample at $3.6$ $\mu m$. The $3.6$ $\mu m$ PL relation for the LMC \citep{Ngeow2008} is also plotted with dotted lines representing  $3\sigma$ deviation. For the model LMC Cepheids  with the canonical mass--luminosity relation, there are two regimes where the luminosity is  $3\sigma$ larger than the luminosity from the observed PL relation.  There are two Cepheids at a period of about $15$ days  and the two near four days having the maximum possible mass--loss rate with a luminosity four magnitudes larger than that given by the PL relation. The models with convective overshooting, shown in the Right Panel of Figure \ref{f7}, show a similar behavior at $15$ days, but at shorter periods the luminosity of the LMC Cepheids are consistent with the PL relation.  

In the analysis of \cite{Ngeow2008}, a number of Cepheids are removed from the determination of the infrared PL relation because the IR brightness is too large. This is presumably due to blending, however infrared excess due to mass loss has the same effect.  The predicted luminosity of the model Cepheids is consistent with the range of luminosities of these outliers, providing possible observational agreement.

The computation of mass--loss rates of the theoretical Cepheid models predicts that at some points on the second crossing of the instability strip the mass--loss rates are increased by up to three orders of magnitude.  This significant increase agrees with the result in the section 2.

\section{Mass Loss and the Mass Discrepancy}

It has been shown that mass loss increases as metallicity decreases for theoretical models of Cepheids. This result is consistent with the observations of mass discrepancy as a function of metallicity \citep{Keller2006, Keller2008}.  Furthermore the dependence of mass loss on metallicity is found to change for larger periods where the mass--loss enhancement becomes less significant.  Along with the fact that more massive Cepheids spend less time on the instability strip than less massive Cepheids, this result is potentially consistent with the observation that the mass discrepancy is constant \citep{Keller2008} or decreases \citep{Caputo2005} as a function of mass.  The model of mass loss being driven by radial pulsation and shocks is consistent with the behavior of mass discrepancy between pulsation and evolution calculations as a function of both mass and metallicity.  But can mass loss predict the actual measured differences of mass?

 For mass loss to be the solution it must be able to account for a difference of approximately $1M_\odot$ for Cepheids with an evolutionary mass of $5M_\odot$ and $0$ -- $2M_\odot$ for Cepheids with an evolutionary mass $11M_\odot$  in the Milky Way. The lower limit represents the mass discrepancy with the dependence on mass that is predicted by \cite{Caputo2005} and the upper limit is due to the constant mass discrepancy argued by \cite{Keller2008}.  For LMC Cepheids the mass difference is approximately $1$ -- $1.5M_\odot$ for $5M_\odot$ Cepheids and $0$ -- $2.5M_\odot$ for $11M_\odot$ Cepheids, while for SMC Cepheids the mass difference is $1$ -- $2M_\odot$ and $0$ -- $3M_\odot$ respectively.

A timescale is required to determine the average mass--loss rates needed to account for mass discrepancy.  This timescale is assumed to be the evolutionary lifetime on the second crossing: about $25$ Myr for a $5M_\odot$ Galactic Cepheid and about $2$ Myr for an $11M_\odot$  Galactic Cepheid \citep{Bono2006}.   Furthermore, it is assumed that the evolutionary timescales are similar in the LMC and SMC.  In that case, a $5M_\odot$ Cepheid would need to lose mass at an average rate of $4\times 10^{-8}$ in the Galaxy, $6\times10^{-8}$ in the LMC and $8\times 10^{-8}M_\odot /yr$ in the SMC, to account for the maximum mass difference. For an $11 M_\odot$ Cepheid, the average mass--loss rate needs to be $0$ -- $1\times 10^{-6}$, $0$ -- $1.2\times 10^{-6}$ and $0$ -- $1.5\times10^{-6}M_\odot  /yr$.  According to the theory of radiative--driven mass loss, which dominates over the pulsation effects at long period, the mass--loss rate is of order $10^{-8}M_\odot$, far less than the maximum value of the average mass--loss rate that is required. The mass--loss rate for $l$ Car has been estimated to be  $<2\times 10^{-8} M_\odot  /yr$ \citep{Bohm-Vitense1994}, and for RS Pup the rate is estimated to be $<3.5\times 10^{-6}M_\odot  /yr$ \citep{Deasy1988}.  

The challenge for explaining the mass discrepancy with mass loss is thus the large--mass regime.  However, this challenge disappears if the mass discrepancy is a decreasing function of mass, as found by \cite{Caputo2005}. The mass discrepancy does not need to be zero at large mass, but it must be smaller than the $17\%$ argued by \cite{Keller2008} to agree with predictions from pulsation--driven mass loss.  If the mass discrepancy is smaller for large mass Cepheids, then an average mass--loss rate of the order $10^{-8}M_\odot  /yr$ may resolve the difference.

At smaller mass, $M < 9M_\odot$,  the theory of pulsation--driven mass loss predicts mass--loss rates of order $10^{-8}M_\odot  /yr$ consistent with the average required mass--loss rate for the mass discrepancy. It is not obvious, however, that the average mass--loss rate for  Cepheids over the first and second crossing is this large.  The predicted mass--loss rates of the model Galactic Cepheids from \cite{Bono2000} is generally too small, but there are a number of examples of significant mass loss in short period Cepheids in Paper I.  

In the LMC and SMC the mass discrepancy at smaller mass is  $17$ -- $25\%$, and the required average mass--loss rate to explain this is $6$ -- $8\times 10^{-8}M_\odot  /yr$. At $10$ -- $20$ day periods the mass--loss rates of the theoretical Cepheid models is greater or equal to this required average, as shown in Figure \ref{f3},  but only for two of the points of on the evolutionary track of the LMC and SMC Cepheid.  For the shorter period LMC/SMC Cepheids the mass loss was shown to be unstable for some of the models, but this was solved by using slightly larger values for the period of pulsation for a maximum value of the mass--loss rate.  These maximum values are of the order $10^{-7}M_\odot$ implying that it is possible to account for the required amount of mass loss.

It is worth noting that pulsation--driven mass loss predicts that the mass loss decreases with metallicity at long periods.  If mass loss is the solution to the mass discrepancy then the mass discrepancy must decrease as a function of mass at a larger rate for the LMC and SMC than for the Milky Way, against the assertion of \cite{Keller2008} that the mass discrepancy is constant with mass.  The mass discrepancy may be constant for periods up to 20 days where the mass loss is dominated by pulsation and shocks.  At longer periods, the mass loss depends primarily on radiative driving, which decreases as $(Z/Z_\odot)^2$, thus implying a smaller mass discrepancy in the SMC and LMC compared to the Milky Way.

The results of this work imply it is possible for mass loss to account for the mass discrepancy but the evidence is circumstantial.  However, mass loss accounts for at least some of the mass discrepancy according to this theory.  In that case, it could be argued the remaining mass discrepancy is due to convective core overshoot which in turn affects mass loss.  The most significant effect of convective core overshoot on the global parameters is the ratio of the luminosity and mass, where this ratio is larger for more overshoot.  This larger ratio causes a larger contribution to the acceleration of the wind via the continuum opacity which in turn increases the mass--loss rate of all Cepheids.  By comparing the mass--loss rates in Figures \ref{f3} and \ref{f4}, it is reasonable to argue that mass loss is important even if the helium core is larger due to convective core overshoot in Cepheid progenitors.  The two possible explanations may be differentiated by determining the mass discrepancy for long period ($P > 30d$), massive Cepheids; if the discrepancy is smaller at large masses then mass loss is a probable solution.

\section{Conclusions}

In this article, we have investigated the hypothesis that mass loss is important for the lower metallicity Cepheids in the LMC and SMC. This has been done by rewriting the analytic method for computing pulsation--driven mass--loss rates as a ratio of mass--loss rates for two Cepheids that are alike in effective temperature and luminosity but with different metallicities, and allowing the other parameters to vary, such as the radius and the period of pulsation.  This leads to a parameter space that can be tested to understand the potential behavior of Cepheids at lower metallicity.  Over the majority of the parameter space the relative mass--loss of the two Cepheids is less than unity, reflecting the explicit dependence on metallicity.  However, there exists two regimes where the relative mass--loss may be larger for lower metallicity.  The first regime is where the relative mass is less than unity and the relative radius is larger than unity, and thus the ratio of the pulsation periods may be greater than unity.  The other regime is where the pulsation period of the lower metallicity Cepheid is less than that of the Cepheid with solar metallicity and the ratios of the mass and radius are approximately unity.  There is an additional parameter constraining this result.  The mass loss is only larger for the lower metallicity Cepheid if the mass loss is significant for the similar solar--metallicity Cepheid. In other words, if mass loss is significantly amplified by pulsation for the solar metallicity Cepheid then the mass loss may be further amplified by pulsation for the lower metallicity Cepheid.

We tested if the parameter space that caused the pulsation amplification of mass loss is consistent with with properties of LMC and SMC Cepheids.  This was done using the metallicity correction of the Period--Luminosity relation to determine periods of LMC and SMC Cepheids relative to Galactic Cepheids with the same luminosity.  Also we used the Period--Radius relation to determine how the radius changes for the lower metallicity Cepheids.  The metallicity correction suggests that lower metallicity Cepheids have longer pulsations periods than Galactic Cepheids and the longer pulsation periods imply that the radii of lower metallicity Cepheids are larger than Galactic Cepheids for the same luminosity.  These behaviors are consistent with mass--loss rates being further amplified implying mass loss is important for LMC and SMC Cepheids.

We have calculated mass--loss rates for theoretical models of Large, and Small Magellanic Cloud, and Galactic Cepheids using the method presented in Paper I. It should be noted that the theoretical models use stellar evolution calculations that do not include mass loss in earlier stages of evolution.  This affects the Cepheid mass--luminosity relation.  It is not clear how the pulsation--driven mass--loss rates of massive ($M>10M_\odot$) Cepheids would change if we use the non--linear mass--luminosity relation of \cite{Keller2008}.  The non--linear mass--luminosity relation suggests that at large mass the luminosity is less for a given mass than what is predicted by \cite{Caputo2005}.  This would suggest that pulsation--driven mass--loss rate would be less, but it is unclear what the pulsation period would be in this case.  The pulsation period of a Cepheid following the non--linear mass--luminosity relation may be smaller than the period of a Cepheid following the linear mass--luminosity relation.  If the period is smaller then the shocks that help drive mass--loss are more efficient and the mass--loss rate may increase.  It would be very interesting to test this possibility.

In the calculation of the mass--loss rates, it was found that the peak mass--loss rates in the second crossing of the instability strip in the SMC is almost a factor of $10$ larger than in the LMC, which in turn, is almost $10$ times larger than that in the Milky Way. Furthermore it was found that the period of about $10$--$20$ days is the point where the dependence of mass loss on metallicity changes sign. This result may be interpreted as the mass loss being amplified significantly for the shorter period Cepheids and radiative driving being the dominant driving mechanism in longer period Cepheids.  The larger mass--loss rates for lower metallicity Cepheids is consistent with the conclusion that mass loss may be amplified more by pulsation at lower metallicity. 

It is believed that dust forms in the wind of a Cepheid, generating an infrared excess.  From the predicted mass--loss rates of theoretical models of Cepheids, we compute Color Magnitude Diagrams as an observational prediction.  The result, however, depends on both the dust--to--gas ratio and the mass--loss rate.  The color excess computed here is thus the maximum value because the dust--to--gas ratio used in this work is the maximum value.  The dust--to--gas ratio may be a few times to an order of magnitude less, implying the color excess may be about a factor of 2--3 less.  The predicted $3.6$ $\mu m$ luminosity computed from the sum of the stellar and circumstellar shell luminosities was found to be  consistent with the observed  infrared Period--Luminosity relation at $3.6$ $\mu m$ \citep{Ngeow2008, Freedman2008} as well as possibly agreeing with the set of Cepheids from \cite{Ngeow2008} that were considered outliers with much larger luminosities due to blending effects.

The ability of mass loss to explain the mass discrepancy has also been explored.  It is shown that pulsation--driven mass loss is a possible explanation, but there still is not enough information to say this with certainty.   The most difficult aspect to explain is the $17\%$ mass discrepancy at large mass where the predicted mass--loss rates are smaller than the necessary values of $10^{-6}M_\odot  /yr$. However, the $17\%$ mass discrepancy is an upper limit and the mass discrepancy at large masses range from $17\%$ down to zero.  At smaller masses the pulsation--driven mass loss is consistent with a mass discrepancy of $17\%$.  For the LMC and SMC the mass--loss rates increase,  implying mass loss can also explain the metallicity dependence of the mass discrepancy.

The results of this work suggest that it is possible for mass--loss rates of lower metallicity Cepheids in the  LMC and SMC to be greater than the mass--loss rates of similar Galactic Cepheids. This result is counterintuitive, one might expect mass loss to scale as the metallicity.  However, the pulsation period, radius and other parameters that describe a Cepheid and its mass loss also have dependencies on the metallicity, and the combination of these parameters and the dependencies cause the mass--loss rates to be potentially more significant.  Furthermore, the amount of mass loss predicted for Magellanic Cloud Cepheids is consistent with the observed behavior of the mass discrepancy of Cepheids as a function of metallicity from \cite{Keller2006} as well as the dependence on mass \cite{Caputo2005}.  Therefore mass loss cannot be discounted as a possible contribution to the mass discrepancy.  The amount of mass loss calculated using the pulsation mass--loss model is significant but there is not enough information to argue mass loss alone resolves the mass discrepancy.    This needs to be tested by with a large number of nonlinear pulsation models of Cepheids on both the first and second crossing to account for the effect of mass loss of the evolution of the period.

\begin{acknowledgements}
The authors appreciate the helpful comments from the referee. HRN is grateful for funding from the Walter John Helm OGSST and the Walter C. Sumner Memorial Fellowship.
\end{acknowledgements}

\bibliography{wind_th}
\bibliographystyle{apj} 
\end{document}